\documentclass[%
reprint,
 showpacs,
 amsmath,amssymb,
]{revtex4-1}
\usepackage{graphicx}
\usepackage{dcolumn}
\usepackage{bm}

\usepackage{CJK}

\usepackage{color}
\usepackage{amssymb}
\usepackage{amsfonts}

\definecolor{Blue}{rgb}{0,0,1}
\definecolor{Bluish}{rgb}{0.5,0,1}

\usepackage[colorlinks,urlcolor=blue,linkcolor=blue,citecolor=blue,anchorcolor=blue]{hyperref}
\makeatletter

\newcommand{\Rmnum}[1]{\expandafter\@slowromancap\romannumeral #1@}
\makeatother
\begin{document}

\preprint{APS/123-QED}

\title{Diverse Entanglement Mechanisms in Multimode Nonlinear Continuous Variables}
\author{Da Zhang$^1$}
\author{David Barral$^{2,3}$}
\author{Yanpeng Zhang$^4$}
\author{Kamel Bencheikh$^5$}
\affiliation{%
$^1$\mbox{School of Physics and Information Engineering, Shanxi Normal University, Taiyuan 030031, China} \\
$^2$\mbox{Laboratoire Kastler Brossel, Sorbonne Universit$\acute{e}$, CNRS, ENS-PSL Research University, Coll\`ege de France,}
\mbox{4 place Jussieu, F-75252 Paris, France}
 \\
$^3$\mbox{Centro de Supercomputaci$\acute{o}$n de Galicia (CESGA), 15705, Santiago de Compostela, Spain } \\
$^4$\mbox{Key Laboratory for Physical Electronics and Devices of the Ministry of Education \& Shaanxi Key Lab of Information }
\mbox{Photonic Technique, School of Electronic and Information Engineering, Xi'an Jiaotong University, Xi'an 710049, China} \\
$^5$\mbox{Centre de Nanosciences et de Nanotechnologies, CNRS, Universit$\acute{e}$ Paris-Saclay, 91120 Palaiseau, France}\\
}%
\date{\today}

\begin{abstract}
\noindent

Non-Gaussian entangled states play a crucial role in harnessing quantum advantage in continuous-variable quantum information. However, how to fully characterize \emph{N}-partite ($N>3$) non-Gaussian entanglement without quantum state tomography remains elusive, leading to a very limited understanding of the underlying entanglement mechanism.
Here, we propose several necessary and sufficient conditions for the positive-partial-transposition separability of multimode nonlinear quantum states resulting from high-order Hamiltonians and successive beam splitting operations.
When applied to the initial state, the beam-splitter operations induce the emergence of different types of entanglement mechanisms, including pairwise high-order entanglement, collective high-order entanglement and the crossover between the two.
We show numerically that for the four-mode scenario, the threshold for the existence of entanglement for any bipartition does not exceed the entanglement of the original state at fixed high-order moments.
These results provide a new perspective for understanding multipartite nonlinear entanglement and will promote their application in quantum information processing.
\end{abstract}

\maketitle
Entanglement, a phenomenon sometimes referred to as "spooky action at a distance", was proposed by Einstein, Podolsky and Rosen (EPR) in a \emph{gedanken} experiment to address the incompleteness of quantum mechanics \cite{einstein.pr.47.777.1935}.
Although the experimental verification of Bell's inequality ultimately refuted EPR's local hidden-variable arguments \cite{clauser.prl.28.938.1972,aspect.prl.49.91.1982}, EPR triggered one of the most fruitful fields of physics: quantum information science \cite{bennett2000quantum}.
Nowadays, it is widely acknowledged that multipartite entanglement --entanglement among $N$ parties-- is considered a fundamental resource for quantum communication \cite{furasawa.science.282.706.1998,pan.nature.518.2015}, quantum cryptography \cite{Gisin.rmp.74.145.2002}, quantum metrology \cite{Vittorio.PRL.96.010401.2006,lu.prl.129.070502.2022}, and quantum computing \cite{walther.nat.434.169.2005,eisert.pra.85.062318.2012}.
Thus, improving our knowledge of its nature is not only crucial for understanding the underlying theory of quantum mechanics, but can also stimulate the generation of new entanglement-based quantum information protocols.

Analogous to the position and the  momentum of the particles addressed in the original EPR paper \cite{einstein.pr.47.777.1935}, the entangled quadratures of electromagnetic fields are commonly generated by bilinear Hamiltonians \cite{Heidmann.prl.59.2555.1987,zhangda.pra.96.043847.2017}.
The resulting twin-photon states are normally distributed in quadratures --thus dubbed Gaussian states \cite{RevModPhys.84.621}.
Starting from quadratic Hamiltonians, multimode Gaussian states have been prepared using different methods, such as linear networks consisting of beam-splitter operations \cite{su.prl.98.070502.2007,Su:12,Yukawa.pra.78.012301.2008} or multiplexing \cite{Pysher.prl.107.030505.2011,armstrong.nc.3.2033.2012,Pfister.prl.112.120505.2014,Roslund2014,jing.prl.125.140501.2020,Furusawa.science.366.2019}.
The positive partial transpose (PPT) criterion \cite{simon.prl.84.2726.2000,werner.prl.86.3658.2001}, the van Loock-Furusawa inequalities \cite{vanlook.pra.67.052315.2003} and the entanglement witnesses proposed in \cite{Sperling.pra.111.110503.2013,Gerke.prl.114.050501.2015,Gerke.prl.117.110502.2016} have achieved great success in characterizing the multipartite entanglement of these states.
However, quantum information tasks based on Gaussian statistics can be efficiently simulated with a classical computer \cite{bartlett.prl.88.097904.2002}.
Non-Gaussian states are thus an essential resource for quantum advantage in quantum information tasks.

Non-Gaussian entangled states carry statistical moments of quadratures beyond Gaussian.
They have been proved to be a necessary resource for bosonic quantum-computational advantage \cite{Mattia.prl.130.090602.2023}, continuous-variable entanglement distillation \cite{masahide.np.4.178.2010,Jaromir.prl.89.137904.2002}, quantum sensing \cite{augusto.science.345.2014} and quantum imaging \cite{liu.prapplied.16.064037.2021}. Multimode non-Gaussian entangled states are usually prepared by applying local non-Gaussian operations to Gaussian states \cite{ourjoumtsev.np.5.189.2009,ra.np.non.2020}.
However, these operations such as photon addition or subtraction are probabilistic.
In order to preserve the key advantage of the continuous-variable regime --determinism, the unconditional preparation of non-Gaussian triple-photon states attracted widespread interest over the past few years \cite{douady.23.2794.ol.2004,Moebius.oe.9.9932.2016,cavanna.pra.101.033840.2020,kangkang.aqt.35.2020,zhang.pra.013704.2021}.
Recently, triple-photon states have been created in a superconducting cavity \cite{chang.prx.10.011011.2020} and their entanglement properties --competition and coexistence of entanglement related to 3rd-order moments and its multiples-- have been characterized \cite{zhang.prl.127.150502.2021,zhang.PRL.130.093602.2023,agust.prl.125.020502,tian.prapplied.18.024065.2022}.

This resource is by construction shared by two or three parties, being thus of limited application in quantum networks. The use of beam-splitter operations or multiplexing techniques on triple-photon states is a novel approach for preparing multimode non-Gaussian states and thus distributing non-Gaussian entanglement.
We refer to the resulting states as multimode nonlinear quantum states, distinct from those produced by non-Gaussian operations.
However, the well-developed criteria based on 2nd-order moments \cite{werner.prl.86.3658.2001,vanlook.pra.67.052315.2003,teh.pra.90.062337.2014} do not capture the nonlinear entanglement.
More importantly, at present, the separability criteria involving high-order moments are mainly limited to two- or three-mode systems \cite{agarwal.njp.211.2005,mm.prl.96.050503.2006,nm.prl.101.130402.2008,walborn.prl.103.160505.2009,walborn.pra.83.032307.2011,Nha.prl.108.030503.2012,
ryo.pra.85.062307.2012,mm.prl.96.050503.2006,nm.prl.101.130402.2008,shchukin.pra.93.032114.2016,zhang.prl.127.150502.2021,zhang.PRL.130.093602.2023,
agust.prl.125.020502,tian.prapplied.18.024065.2022}.
How to diagnose arbitrary multipartite nonlinear entanglement remains elusive.
Not only does this make them poorly understood, it will ultimately be an important experimental concern.

In this work, we propose several necessary and sufficient conditions for multipartite non-PPT entanglement applicable to multimode nonlinear continuous variables.
When applied to a nonlinear entangled state generated by a high-order Hamiltonian, beam-splitter operations can not only increase the number of modes in the system, but also induce the birth of different entanglement mechanisms.
To characterize these different classes of entanglement, multiple vectors are needed to construct the corresponding higher-order covariance matrices (HOCMs).
We introduce new criteria based on the PPT of these HOCMs.
By numerical simulations, we show that a four-mode nonlinear quantum state possesses three different entanglement mechanisms: pairwise high-order entanglement, collective high-order entanglement, and the transition between the two.
The threshold for the existence of these different types of entanglement at fixed high-order moments does not exceed the initial threshold of the states after the nonlinear interaction.
Our results provide a systematic framework for characterizing multipartite nonlinear entanglement and will facilitate its application in quantum information processing.

\begin{figure}[t]
\centering
  \includegraphics[width=7cm]{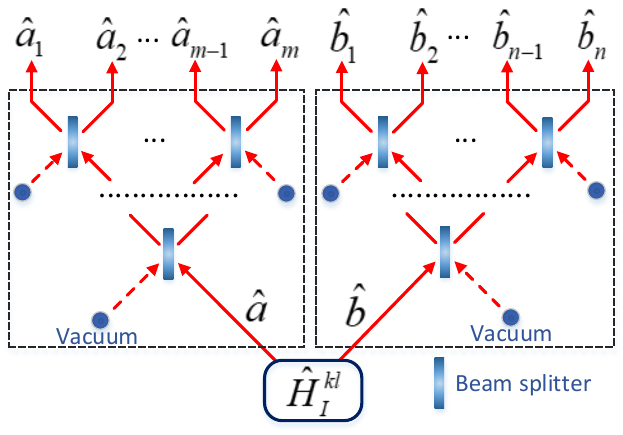}  
  \caption{Sketch of a network of beam splitters used to prepare multimode nonlinear quantum states from a triple-photon state.}
  \label{fig1}
\end{figure}

We begin to study multipartite nonlinear entanglement by considering a variety of multimode quantum states prepared using two arbitrary linear optical networks consisting of beam splitters as shown in Fig. \ref{fig1}.
Without loss of generality, we assume that the initial two-mode nonlinear quantum state described by a density matrix $\hat{\rho}$ is generated by a partially degenerate Hamiltonian $\hat{H}^{kl}_I=i\hbar\kappa\hat{a}^{\dag k}\hat{b}^{\dag l}\hat{p}+H.c$, where the annihilation operators $\hat{a}$, $\hat{b}$ and $\hat{p}$ describe down-conversion modes and pump modes, respectively.
The generated mode $\hat{a}$ ($\hat{b}$) is then mixed with multiple vacuum modes in a linear network.
We label the outputs of the two networks as $\hat{a}_1,\cdots, \hat{a}_m$ and $\hat{b}_1,\cdots, \hat{b}_n$.

A standard approach to characterize multipartite entanglement is to check the separability of all bipartitions.
Our system has $2^{m+n-1}-1$ bipartitions.
Here, for ease of description, we only consider the case where Alice holds the modes $\hat{a}_1$, $\hat{a}_2$, $\cdots,$ and $\hat{a}_m$ and Bob holds $\hat{b}_1, \hat{b}_2, \cdots,$ and $\hat{b}_n$, but the following derivation applies to any bipartition.
We define the high-order quadrature operators $\hat{Q}^{sf}_{\mathcal{O}_i}=[(\hat{o}_i^{f})^s$ $+(\hat{o}_i^{\dag f})^s]/2$ and $\hat{P}^{sf}_{\mathcal{O}_i}=i[(\hat{o}_i^{\dag f})^s-(\hat{o}_i^{f})^s]/2$ ($\mathcal{O}=\mathcal{A}, \mathcal{B}$, $o=a, b$ and $f=k,l$), where $s$ is a positive integer.
For brevity, we now take $s=1$, but the following derivations are general for any $s$.
The vector $\hat{R}^{kl}=(\hat{Q}^k_{\mathcal{A}_1},\hat{P}^k_{\mathcal{A}_1},\cdots,\hat{P}^k_{\mathcal{A}_m},\hat{Q}^l_{\mathcal{B}_1},\cdots,\hat{Q}^l_{\mathcal{B}_n},\hat{P}^l_{\mathcal{B}_n})^T$, grouped together by high-order quadrature operators, allows us to express the generalized commutation relations as
\begin{align}\label{eq1}
[\hat{R}^{kl}_i,\hat{R}^{kl}_j]=i\Omega^{kl}_{ij},
\end{align}
where $\Omega^{kl}=\mathrm{diag}(\Omega^k_\mathcal{A},\Omega^l_\mathcal{B})$. $\Omega^k_\mathcal{A}$ and $\Omega^l_\mathcal{B}$ are multimode matrices corresponding respectively to Alice and Bob and defined in the Supplementary Material \cite{triple.prl.2020}.

Analogously to a Gaussian system, we start our analysis by constructing the multimode HOCMs $V^{kl}$, whose elements are defined as $V^{kl}_{ij}=\langle \hat{R}^{kl}_i \hat{R}^{kl}_j + \hat{R}^{kl}_j \hat{R}^{kl}_i\rangle/2 - \langle\hat{R}^{kl}_i\rangle \langle\hat{R}^{kl}_j\rangle$.
By inserting the commutation relation (\ref{eq1}) and the property $\langle\hat{R}^{kl}_i\rangle=0$ \cite{triple.prl.2020}, we obtain the following uncertainty principle
\begin{align}\label{eq2}
V^{kl}+\frac{i}{2}\langle\Omega^{kl}\rangle \geq 0.
\end{align}
All physical states that satisfy $\langle\hat{R}^{kl}_i\rangle=0$ must obey this inequality.
Inequality (\ref{eq2}) ensures the definite positivity of $V^{kl}$ due to the skew symmetry of $\langle\Omega^{kl}\rangle$.

Let us now introduce necessary and sufficient conditions for separability of HOCMs. The necessary condition for separability is that the partially transposed $\hat{\rho}^{PT}$ is semi-positive definite.
The PPT criterion has been widely used in experiments to diagnose entanglement, especially in the continuous-variable regime \cite{jing.prl.125.140501.2020,jing.prl.124.090501.2020}, as the effect of a partial transpose operation on a CM is easy to describe.
As far as the HOCMs are concerned, one can obtain the partially transposed state for Bob's subsystem by just changing the signs of the $n$ momenta \{$\hat{P}^l_{\mathcal{B}_j}$\} belonging to subsystem $\mathcal{B}$, i.e., $\widetilde{V}^{kl}=TV^{kl}T$, where the matrix $T$ leaves unchanged Alice's modes and performs a mirror reflection on Bob's modes.
Thus, if the state represented by $V^{kl}$ is separable, the partially transposed $\widetilde{V}^{kl}$ still satisfies the  uncertainty principle in the form of
\begin{align}\label{eq3}
\widetilde{V}^{kl}+\frac{i}{2}\langle\Omega^{kl}\rangle \geq 0.
\end{align}
The multimode PPT criterion given by Eq. (\ref{eq3}) can be rewritten in block form as
\begin{align}\label{eq4}
\left(
    \begin{array}{cc}
      A & C \\
      C^T & B \\
    \end{array}
  \right)+
  \frac{i}{2}\left(
    \begin{array}{cc}
      \langle\Omega_\mathcal{A}^k\rangle & 0 \\
      0 & -\langle\Omega^l_\mathcal{B}\rangle \\
    \end{array}
  \right)\geq 0,
\end{align}
where $A$ and $B$ are respectively the local HOCMs of Alice and Bob subsystems, and $C$ describes the correlation between them.

Recently, the PPT criterion was shown to be a sufficient condition for separability in two-mode scenarios \cite{zhang.prl.127.150502.2021}.
However, this sufficiency cannot be extended to general multimode settings.
This is due to the existence of bound entangled states \cite{werner.prl.86.3658.2001,Horodecki.prl.85.2657.2000,Horodecki.2003.Bound}, which are inseparable but with PPT, and therefore not distillable.
However, the PPT criterion is still sufficient for separability under some restrictive conditions, such as Alice's system comprising just one mode $m=1$, or subsystems satisfying local exchange symmetry --bisymmetric states \cite{werner.prl.86.3658.2001,Alessio.pra.71.032349.2005}. Below, we demonstrate through theorems 2-3 and 4 that sufficiency holds for HOCMs in the respective cases.

A bipartite state is said to be separable iff it can be decomposed into a convex mixture of product states.
By definition, the HOCMs of a product state are block diagonal.
Then we have the following theorem concerning the separability of $V^{kl}$.

\emph{Theorem 1.--}A state specified by $V^{kl}$ is separable iff
\begin{align}\label{eq5}
V^{kl} \geq \sigma_\mathcal{A}\oplus \sigma_\mathcal{B}
\end{align}
is established, where $\sigma_\mathcal{A}$ and $\sigma_\mathcal{B}$ are local HOCMs that satisfy the uncertainty principle $\sigma_\mathcal{A}+ i\langle\Omega^k_\mathcal{A}\rangle/2\geq 0$ and $\sigma_\mathcal{B}+ i\langle\Omega^l_\mathcal{B}\rangle/2\geq 0$, respectively.

The proofs of the theorems are in the Supplementary Material \cite{triple.prl.2020}.
Theorem 1 shows that if inequality (\ref{eq5}) is satisfied, then the state characterized by $V^{kl}$ can be obtained from the product state with HOCM $\sigma_\mathcal{A}\oplus \sigma_\mathcal{B}$ by local operations and classical communication.
In practice, condition (\ref{eq5}) is not as accessible as condition (\ref{eq4}), since we have to prove whether $\sigma_\mathcal{A}$ and $\sigma_\mathcal{B}$ exist \cite{giedke.prl.87.167904.2001}.
However, it is useful for studying multipartite nonlinear entanglement.
With respect to the experimentally accessible separability criterion, we have the following theorem for $1\times n$ separability:

\emph{Theorem 2.--}Let $V^{kl}$ be a HOCM in which subsystem $\mathcal{A}$ consists of 1 mode and $n$ modes comprise subsystem $\mathcal{B}$.
$V^{kl}$ is separable iff the inequality (\ref{eq4}) holds.

The separability condition (\ref{eq5}) reveals that for $\sigma_\mathcal{A}$ and $\sigma_\mathcal{B}$ satisfying the local uncertainty principle, $V^{kl}$ is a subset of the set of separable states iff $V^{kl}-(\sigma_\mathcal{A}\oplus0_\mathcal{B})\geq(0_\mathcal{A}\oplus\sigma_\mathcal{B})$.
From the positivity and variational characterisations of the Schur complement \cite{zhang2006schur,horn2012matrix}, the necessary and sufficient condition of separability is equivalent to the existence of a 2$\times$2 real matrix $\sigma_\mathcal{A}$ such that $A-C\left(B-i\langle\Omega^l_\mathcal{B}\rangle/2\right)^{-1}C^T\geq\sigma_\mathcal{A}\geq -i\langle\Omega^k_\mathcal{A}\rangle/2$, which is the upper and lower bounds given by inequalities (\ref{eq2}) and (\ref{eq4}).
The existence of $\sigma_\mathcal{A}$ has been proved in \cite{lami2018gaussian}, thus confirming the sufficient necessity of the PPT criterion.

The HOCM $A$ constructed from the vector $\hat{R}^{k}=(\hat{Q}^k_{\mathcal{A}_i},$ $\hat{P}^k_{\mathcal{A}_i})^T$ describes only partial information of modes $\hat{a}_i$.
However, we can define multimode quadrature operators $\hat{Q}^{k_1\cdots k_m}_{\hat{a}_1\cdots\hat{a}_m}=(\hat{a}_1^{\dag k_1}\cdots\hat{a}_m^{\dag k_m}+\hat{a}_1^{k_1}\cdots\hat{a}_m^{k_m})/2$ and $\hat{P}^{k_1\cdots k_m}_{\hat{a}_1\cdots\hat{a}_m}=i(\hat{a}_1^{\dag k_1}\cdots\hat{a}_m^{\dag k_m}-\hat{a}_1^{k_1}\cdots\hat{a}_m^{k_m})/2$, which satisfy some commutation relations \cite{triple.prl.2020}.
If we replace the vector $\hat{R}^{k}$ by $\hat{R}'^{k}=(\hat{Q}^{k_1\cdots k_m}_{\hat{a}_1\cdots\hat{a}_m},\hat{P}^{k_1\cdots k_m}_{\hat{a}_1\cdots\hat{a}_m})^T$ the new HOCM $A'$ encompasses $m$ modes. Correspondingly, Theorem 2 can be extended to $m \times n$ separability as follows:

\begin{figure*}[htpb]
\centering
  \includegraphics[width=17cm]{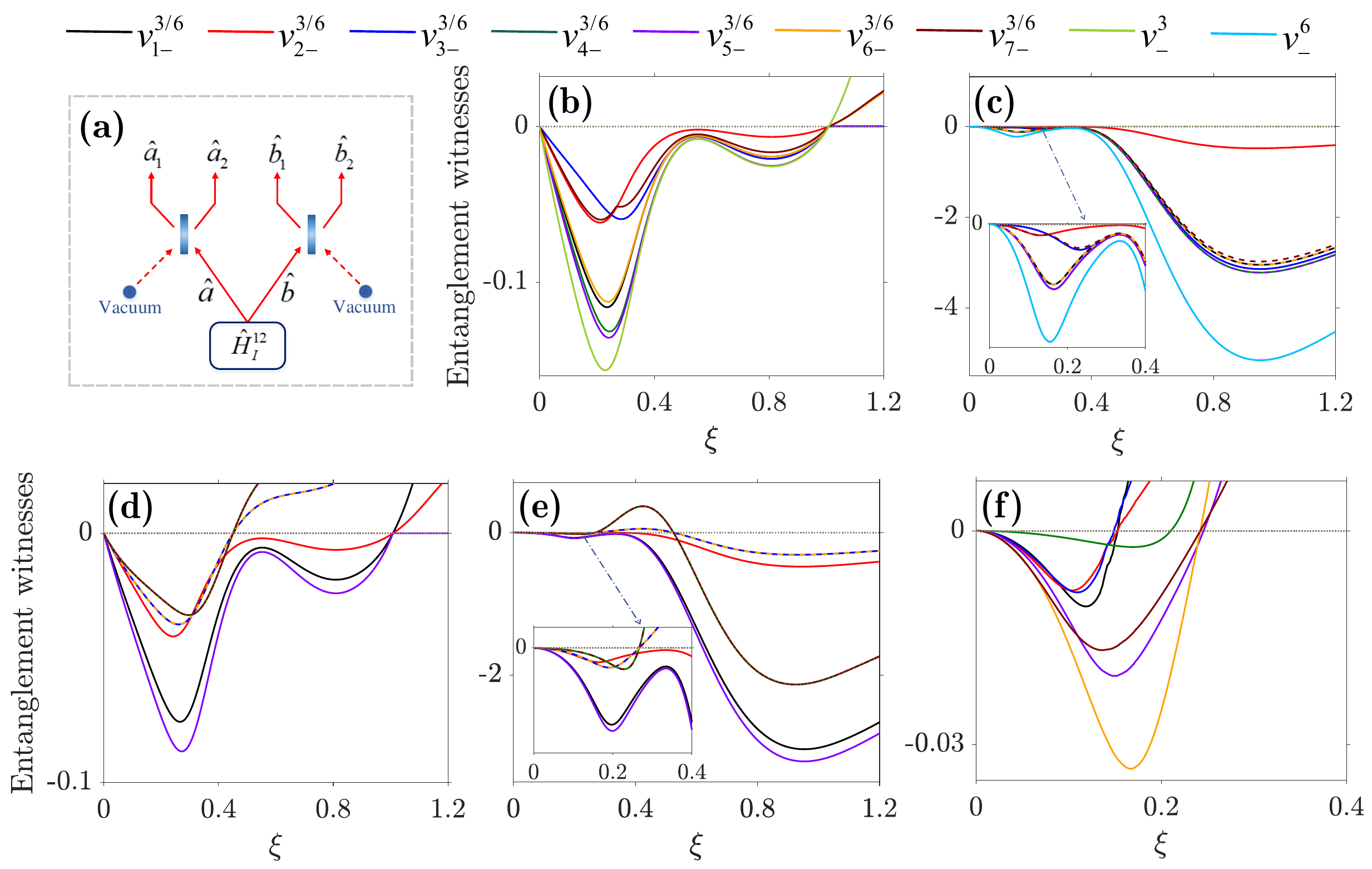}  
  \caption{(a) Schematic diagram for the preparation of a four-mode nonlinear quantum state, in which the transmittance of both beam splitters is $3/4$.
  (b)-(f) Evolution of $\nu_{i-}^{n}$ with the interaction strength $\xi$, where $\alpha_p=\sqrt{25}$. $\nu_{i-}^{n}<0$ implies that the corresponding bipartition is entangled at $n$th-order moment. $\nu_{-}^{3}<0$ ($\nu_{-}^{6}<0$) indicates that the original modes $\hat{a}$ and $\hat{b}$ are entangled at 3rd-order (6th-order) moment.}
  \label{fig2}
\end{figure*}
\emph{Theorem 3.--}Let $V_{\uppercase\expandafter{\romannumeral1}}^{kl}$ be such a HOCM constructed from multimode quadrature operators in which subsystem $\mathcal{A}$ consists of $m$ modes and $n$ modes comprise subsystem $\mathcal{B}$.
$V_{\uppercase\expandafter{\romannumeral1}}^{kl}$ is separable iff the inequality (\ref{eq4}) holds.

Now consider the bisymmetric $V^{kl}$ built from arbitrary vectors.
Bisymmetry implies that $V^{kl}$ is invariant after exchanging any two modes in the subsystem $\mathcal{A}$ ($\mathcal{B}$) \cite{Alessio.pra.71.032349.2005}.
This requires that all modes in each subsystem have the same energy and origin, i.e., they all come from mode $\hat{a}$ ($\hat{b}$).
For such a $V^{kl}$, there always exists a local symplectic transformation acting on the subsystem that transforms $V^{kl}$ into a $(1 \times n)$-mode bipartite state and $m-1$ uncorrelated single-mode states \cite{triple.prl.2020}.
By applying theorem 2 to the transformed $V^{kl}$, the following result is obtained for bisymmetric states:

\emph{Theorem 4.--}PPT criterion is a necessary and sufficient condition for separability of bisymmetric states described by $V^{kl}$.

Theorems 2, 3 and 4 are natural extensions of the PPT criterion \cite{simon.prl.84.2726.2000,werner.prl.86.3658.2001,Alessio.pra.71.032349.2005} to multimode non-Gaussian systems.
These criteria, which rely on different quadrature-basis vectors, unveil disparate entanglement mechanisms. Thus, we demonstrate now the rich entanglement structure possessed by multimode nonlinear states as the one shown in Fig. \ref{fig2}(a). We consider an input state generated by a Hamiltonian $\hat{H}^{12}_I=i\hbar\kappa\hat{a}^{\dag}\hat{b}^{\dag 2}\hat{p}+H.c.$. Then, down-conversion modes $\hat{a}$ and $\hat{b}$ pass  respectively through beam splitters with transmittance of $3/4$.
Violation of criterion (\ref{eq3}) is equivalent to fulfilling the inequality $\nu^{n}_{i-}<0$, where $\nu^{n}_{i-}$ is the minimum eigenvalue of $\widetilde{V}^{kl}+i\langle\Omega^{kl}\rangle/2$, the superscript $n=k+l$ represents the order of the HOCM, and the subscript $i=1,...,7$, corresponding in turn to the seven bipartitions  $\hat{a}_1|\hat{a}_2\hat{b}_1\hat{b}_2$, $\hat{a}_2|\hat{a}_1\hat{b}_1\hat{b}_2$, $\hat{b}_1|\hat{a}_1\hat{a}_2\hat{b}_2$, $\hat{b}_2|\hat{a}_1\hat{a}_2\hat{b}_1$, $\hat{a}_1\hat{a}_2|\hat{b}_1\hat{b}_2$, $\hat{a}_1\hat{b}_1|\hat{a}_2\hat{b}_2$, and $\hat{a}_1\hat{b}_2|\hat{a}_2\hat{b}_1$.
Due to the complex entanglement properties of this state, we classify them into the following three categories: pairwise high-order entanglement (Type-\uppercase\expandafter{\romannumeral1}), collective high-order entanglement (Type-\uppercase\expandafter{\romannumeral3}), and the transition between the two (Type-\uppercase\expandafter{\romannumeral2}).

\emph{Type-\uppercase\expandafter{\romannumeral1} entanglement}:
The 3rd-order CM $V^{12}$ is constructed with a high-order quadrature vector $\hat{R}^{12}=(\hat{Q}^1_{\mathcal{A}_1},\hat{P}^1_{\mathcal{A}_1},\hat{Q}^1_{\mathcal{A}_2},\hat{P}^1_{\mathcal{A}_2},
\hat{Q}^2_{\mathcal{B}_1},\hat{P}^2_{\mathcal{B}_1},\hat{Q}^2_{\mathcal{B}_2},$ $\hat{P}^2_{\mathcal{B}_2})^T$.
Figure \ref{fig2}(b) shows the evolution of $\nu^{3}_{i-}$ with the interaction strength $\xi=\kappa t\alpha_p$, where $\alpha_p$ is the amplitude of pump field and $t$ is the time of the system evolution.
Interestingly, all the 7 bipartitions are entangled for $\xi<1$.
Note that the entanglement of the original modes $\hat{a}$ and $\hat{b}$ loaded on the 3rd-order CM also disappears when $\xi\geq1$, as shown in Fig. \ref{fig2}(b) (yellowblue line).
This indicates that the threshold for the existence of quadripartite 3rd-order-moment nonlinear entanglement generated by beam-splitter operations does not exceed the initial state.
Furthermore, the 6th-order CM is obtained using now  $s=2$.
Correspondingly, the evolution of $\nu^6_{i-}$ as a function of $\xi$ is presented in Fig. \ref{fig2}(c).
Each bipartition is entangled over the entire parameter range like the original state (deepskyblue line in Fig. \ref{fig2}(c)).
Remarkably, with the increase of $\xi$, the fully inseparable quadripartite entanglement gradually transits from the 3rd-order CM to the 6th-order CM.

\emph{Type-\uppercase\expandafter{\romannumeral2} entanglement}: The 3rd-order CM $V^{12}_{\uppercase\expandafter{\romannumeral1}_1}$ is now constructed with a vector $\hat{R}_{\uppercase\expandafter{\romannumeral1}_1}^{12}=(\hat{Q}^1_{\mathcal{A}_1},\hat{P}^1_{\mathcal{A}_1},\hat{Q}^1_{\mathcal{A}_2},$ $\hat{P}^1_{\mathcal{A}_2},\hat{Q}^{11}_{\mathcal{B}_1\mathcal{B}_2},\hat{P}^{11}_{\mathcal{B}_1\mathcal{B}_2})^T$.
Since modes $\hat{b}_1$ and $\hat{b}_2$ form a \emph{local} mode in $V^{12}_{\uppercase\expandafter{\romannumeral1}_1}$, the corresponding criterion (\ref{eq3}) can only  determine the inseparability of bipartitions $\hat{a}_1|\hat{a}_2\hat{b}_1\hat{b}_2$, $\hat{a}_2|\hat{a}_1\hat{b}_1\hat{b}_2$, and $\hat{a}_1\hat{a}_2|\hat{b}_1\hat{b}_2$, respectively.
To diagnose the separability of the other four bipartitions, we need to construct two 3rd-order CMs $V^{12}_{\uppercase\expandafter{\romannumeral1}_2}$ and $V^{12}_{\uppercase\expandafter{\romannumeral1}_3}$ with $\hat{R}^{12}_{\uppercase\expandafter{\romannumeral1}_2}=(\hat{Q}^1_{\mathcal{A}_1},$ $\hat{P}^1_{\mathcal{A}_1},$ $\hat{Q}^{11}_{\mathcal{A}_2\mathcal{B}_1},
\hat{P}^{11}_{\mathcal{A}_2\mathcal{B}_1},\hat{Q}^1_{\mathcal{B}_2},\hat{P}^1_{\mathcal{B}_2})^T$
and $\hat{R}^{12}_{\uppercase\expandafter{\romannumeral1}_3}=(\hat{Q}^1_{\mathcal{A}_1},\hat{P}^1_{\mathcal{A}_1},$ $\hat{Q}^1_{\mathcal{B}_1},\hat{P}^1_{\mathcal{B}_1},
\hat{Q}^{11}_{\mathcal{A}_2\mathcal{B}_2},\hat{P}^{11}_{\mathcal{A}_2\mathcal{B}_2})^T$ as basis vectors, respectively.
Figure \ref{fig2}(d) shows how the different eigenvalues $\nu^{3}_{i-}$ evolve with $\xi$.
At 3rd-order moments, the threshold for the existence of entanglement for the bipartitions $\hat{a}_1|\hat{a}_2\hat{b}_1\hat{b}_2$, $\hat{a}_2|\hat{a}_1\hat{b}_1\hat{b}_2$, and $\hat{a}_1\hat{a}_2|\hat{b}_1\hat{b}_2$ is $\xi=1$, while for the other four bipartitions is $\xi=0.45$, none of which exceeding the initial state.
Figure \ref{fig2}(e) shows the evolution of the eigenvalues $\nu^6_{i-}$ obtained from 6th-order CMs ($s=2$).
The entanglement of bipartitions $\hat{b}_1|\hat{a}_1\hat{a}_2\hat{b}_2$, $\hat{b}_2|\hat{a}_1\hat{a}_2\hat{b}_1$, $\hat{a}_1\hat{b}_1|\hat{a}_2\hat{b}_2$ and $\hat{a}_1\hat{b}_2|\hat{b}_1\hat{a}_2$ disappears in the range $0.27\leq\xi\leq0.52$, but in other parameter regions the quadripartite system is still fully inseparable.
Other types of entanglement structures at 6th-order moments are provided in \cite{triple.prl.2020}.

\emph{Type-\uppercase\expandafter{\romannumeral3} entanglement}: The 6th-order CM $V^{24}_{\uppercase\expandafter{\romannumeral2}_1}$ created by $\hat{R}_{\uppercase\expandafter{\romannumeral2}_1}^{24}=
(\hat{Q}^{1}_{\mathcal{A}_1},\hat{P}^{1}_{\mathcal{A}_1},\hat{Q}^{113}_{\mathcal{A}_2\mathcal{B}_1\mathcal{B}_2}, $ $\hat{P}^{113}_{\mathcal{A}_2\mathcal{B}_1\mathcal{B}_2})^T$ has only two \emph{local} modes, so the corresponding PPT criterion can only verify the separability of the bipartition $\hat{a}_1|\hat{a}_2\hat{b}_1\hat{b}_2$.
The entanglement conditions for the residual bipartitions are constructed with the vectors
\begin{align*}
 \hat{R}_{\uppercase\expandafter{\romannumeral2}_2}^{24}=
(\hat{Q}^{1}_{\mathcal{A}_2},\hat{P}^{1}_{\mathcal{A}_2},\hat{Q}^{113}_{\mathcal{A}_1\mathcal{B}_1\mathcal{B}_2}, \hat{P}^{113}_{\mathcal{A}_1\mathcal{B}_1\mathcal{B}_2})^T , \\
\hat{R}_{\uppercase\expandafter{\romannumeral2}_3}^{24}=
(\hat{Q}^{1}_{\mathcal{B}_1},\hat{P}^{1}_{\mathcal{B}_1},\hat{Q}^{113}_{\mathcal{A}_1\mathcal{A}_2\mathcal{B}_2},
\hat{P}^{113}_{\mathcal{A}_1\mathcal{A}_2\mathcal{B}_2})^T \\
\hat{R}_{\uppercase\expandafter{\romannumeral2}_4}^{24}=
(\hat{Q}^{3}_{\mathcal{B}_2},\hat{P}^{3}_{\mathcal{B}_2},\hat{Q}^{111}_{\mathcal{A}_1\mathcal{A}_2\mathcal{B}_1}, \hat{P}^{111}_{\mathcal{A}_1\mathcal{A}_2\mathcal{B}_1})^T ,\\
\hat{R}_{\uppercase\expandafter{\romannumeral2}_5}^{24}=
(\hat{Q}^{11}_{\mathcal{A}_1\mathcal{A}_2}, \hat{P}^{11}_{\mathcal{A}_1\mathcal{A}_2},\hat{Q}^{13}_{\mathcal{B}_1\mathcal{B}_2},\hat{P}^{13}_{\mathcal{B}_1\mathcal{B}_2})^T , \\
\hat{R}_{\uppercase\expandafter{\romannumeral2}_6}^{24}=
(\hat{Q}^{11}_{\mathcal{A}_1\mathcal{B}_1}, \hat{P}^{11}_{\mathcal{A}_1\mathcal{B}_1}, \hat{Q}^{13}_{\mathcal{A}_2\mathcal{B}_2},\hat{P}^{13}_{\mathcal{A}_2\mathcal{B}_2})^T,
\end{align*}
and
$\hat{R}_{\uppercase\expandafter{\romannumeral2}_7}^{24}=
(\hat{Q}^{13}_{\mathcal{A}_1\mathcal{B}_2},\hat{P}^{13}_{\mathcal{A}_1\mathcal{B}_2},\hat{Q}^{11}_{\mathcal{B}_1\mathcal{A}_2}, \hat{P}^{11}_{\mathcal{B}_1\mathcal{A}_2})^T$, respectively.
We show the evolution of $\nu^6_{i-}$ with $\xi$ in Fig. \ref{fig2}(f).
Compared with other types of entanglement related to 6th-order moments, this type of fully inseparable quadripartite entanglement exists in a narrower parameter interval.
Notably, this collective high-order entanglement is similar to that of a fully nondegenerate triple-photon state \cite{agust.prl.125.020502,zhang.PRL.130.093602.2023}.
Following the same idea, the collective quadripartite entanglement also exists in the 6th-order CM constructed by the vector $\hat{R}_{\uppercase\expandafter{\romannumeral2}_8}^{24}=
(\hat{Q}^{1}_{\mathcal{A}_1}, \hat{P}^{1}_{\mathcal{A}_1}, \hat{Q}^{131}_{\mathcal{A}_2\mathcal{B}_1\mathcal{B}_2}, \hat{P}^{131}_{\mathcal{A}_2\mathcal{B}_1\mathcal{B}_2})^T$.
The associated numerical verification is given in \cite{triple.prl.2020}.

Zero local moments, i.e. $\langle\hat{R}^{kl}\rangle=0$, is one of the fundamental properties of state generated by high-order Hamiltonians \cite{zhang.prl.127.150502.2021,zhang.PRL.130.093602.2023,hillery.pra.42.498.1992}, and this property is invariant after beam-splitting operations.
Therefore, criterion (\ref{eq4}) is applicable to arbitrary multimode nonlinear quantum states.
By choosing appropriate observables, the different mechanisms of entanglement can be systematically characterized.

In summary, we proposed several sufficient and necessary conditions for the positive-partial-transposition separability of multimode nonlinear quantum states resulting from high-order Hamiltonians and beam-splitter operations.
These criteria present some appealing advantages.
First, they are efficient in terms of resources compared with other state-of-the-art methods based on the H$\ddot{o}$lder inequality \cite{hillery.pra.81.062322.2010}, Fisher information \cite{gessner2017entanglement} or quantum tomography-based methods \cite{wm.prl.119.183601.2017}.
Second, our strategy is platform-independent as it can be applied to any physical system where information is encoded in continuous variables.
Third, the high-order moments involved in our criteria are within reach with current coherent-detection methods \cite{ra.np.non.2020, chang.prx.10.011011.2020}. Finally, they reveal different non-Gaussian entanglement mechanisms. Remarkably, we found that a four-mode nonlinear quantum state possesses three different entanglement mechanisms, which has never been reported in continuous variables. This naturally leads to an interesting question, what is the advantage of nonlinear quantum states with multiple entanglement mechanisms in quantum information protocols?
Our results not only provide a solid basis for the experimental diagnosis of multipartite nonlinear entanglement, but also promote the understanding of the interaction between Gaussian operations and non-Gaussian entangled states.

\section*{Acknowledgement}
We thank Qiongyi He and Yu Xiang for insightful discussions.
This work was supported by the National Natural Science Foundation of China (12204293, 61975159), Applied Basic Research Program in Shanxi Province (202203021212387), and by the Agence Nationale de la Recherche through Project TRIQUI (ANR 17-CE24-0041).


%

\end{document}